\documentclass[preprint,showpacs,preprintnumbers,amsmath,amssymb]{revtex4}

\usepackage{graphicx}% Include figure files
\usepackage{dcolumn}% Align table columns on decimal point
\usepackage{bm}% bold math

\begin{document}

\title{Brane world corrections to scalar vacuum force in RSII-p}

\author{Rom\'an Linares$^1$}
\email{lirr@xanum.uam.mx}
\author{Hugo A. Morales-T\'ecotl$^{1,2}$}
\email{hugo@xanum.uam.mx}
\author{Omar Pedraza$^{1,3}$}
\email{opedrazao@ipn.mx}

\affiliation{$^1$Departamento de F\'{\i}sica, Universidad Aut\'onoma Metropolitana Iztapalapa,\\
San Rafael Atlixco 186, C.P. 09340, M\'exico D.F., M\'exico,}

\affiliation{$^2$ Instituto de Ciencias Nucleares, Universidad Nacional Aut\'onoma de M\'exico,\\
A. Postal 70-543, M\'exico D.F., M\'exico,}

\affiliation{ $^3$Centro de Investigaci\'on en Ciencia Aplicada y Tecnolog\'{\i}a Avanzada,\\
\, Unidad Legaria, Instituto Polit\'ecnico Nacional, \, \\
Av. Legaria 694, C.P. 11500, M\'exico D.F., M\'exico.
}


\begin{abstract}
Vacuum force is an interesting low energy test for brane worlds due to its dependence on field's modes and its role in submillimeter gravity experiments. In this work we generalize a previous model example: the scalar field vacuum force between two parallel plates lying in the brane of a Randall-Sundrum scenario extended by $p$ compact dimensions (RSII-$p$). Upon use of Green's function technique, for the massless scalar field, the 4D force is obtained from a zero mode while corrections turn out attractive and depend on the separation between plates as $l^{-(6+p)}$. For the massive scalar field a quasilocalized mode yields the 4D force with attractive corrections behaving like $l^{-(10+p)}$. Corrections are negligible w.r.t. 4D force for $AdS_{(5+p)}$ radius less than $\sim 10^{-6}$m. Although the $p=0$ case is not physically viable due to the different behavior in regard to localization for the massless scalar and electromagnetic fields it yields an useful comparison between the dimensional regularization and Green's function techniques as we describe in the discussion.
\end{abstract}

%\keywords{Field Theories in Higher Dimensions}%Use showkeys class option if keyword
\maketitle
\section{Introduction}

Spacetimes including more than three spatial dimensions have been
studied since the first attempt of N\"{o}rdstrom to unify
electromagnetism and scalar gravity \cite{Nordstrom:1988fi}, and the
subsequent ones of Kaluza \cite{Kaluza:1921tu} and Klein
\cite{Klein:1926tv} to unify general relativity and
electromagnetism. Later on the idea was put forward that our
observable universe was constrained to a brane subspace of a higher
dimensional spacetime
\cite{Akama:1982jy,Rubakov:1983bb,Visser:1985qm} to try to solve
problems like the cosmological constant one.

More recently,
motivated by M theory \cite{Horava:1996ma,Horava:1995qa}, and
following phenomenological insights like trying to solve the
hierarchy problem such an idea has been revived
\cite{Randall:1999vf,Randall:1999ee,Antoniadis:1998ig,Arkani-Hamed:1998rs,Antoniadis:1990ew}.
In these brane world models, to be consistent with particle physics
and recent submillimeter gravity experiments
\cite{Hoyle:2000cv,Adelberger:2006dh,Kapner:2006si}, the Standard
Model fields are trapped on the brane whereas gravity is spread out
over the full higher dimensional spacetime. Although extra
dimensions were originally considered to be small near the Planck
length \cite{Klein:1926tv}, in brane worlds they could be as large
as say $1 TeV^{-1}$ or even infinite
\cite{Arkani-Hamed:1998rs,Randall:1999ee}. Their effects range from
cosmology to particle physics and astrophysics (see e.g.
\cite{Rubakov:2001kp,
Feruglio:2004zf,Maartens:2003tw,PerezLorenzana:2005iv}).

Low energy tests on the other hand have remained largely unexplored
based on the idea that acceptable brane world scenarios should be
built to be compatible with them. In contraposition it has been
recently advanced that subjecting brane world scenarios to low
energy tests may shed light on their viability. Indeed already
available experimental data can be used in this way. Hence high
precision atomic and Casimir force experiments become relevant
\cite{Bluhm:2000tv,Luo:2006ck,Linares:2007yz,MoralesTecotl:2006eh,Linares:2005cj}.
The vacuum force predicted by H.B.G. Casimir in 1948
\cite{Casimir:1948dh} between neutral perfect conducting plates,
which has been confirmed experimentally
\cite{Sparnaay:1958wg,Mohideen:1998iz,Bressi:2002fr,Decca:2005yk,Klimchitskaya:2005df,Lamoreaux:2005gf}
involves two aspects naturally appearing in the study of brane
worlds, namely the mode structure of matter fields and the
submillimeter length scale, of order $1 \mu$m, for which it becomes
noticeable. In 4D such a force has been understood by means of QED
\cite{Milton:2001yy} and the modification of the modes due to the
presence of the plates. Extensions to other fields, geometries,
materials, boundaries and models have been also studied
\cite{Milton:2001yy}. Regarding extra dimensions the Casimir force
has been discussed in string theory
\cite{Fabinger:2000jd,Gies:2003cv,Brevik:2000fs,Hadasz:1999tr} and
Randall-Sundrum models
\cite{Elizalde:2002dd,Garriga:2002vf,Pujolas:2001um,Flachi:2001pq,Goldberger:2000dv}
as well as within inflationary brane world universe models
\cite{Nojiri:2000bz} and dark matter
\cite{Durrer:2007ww,Ruser:2007wz}.

Notice that whereas in 4D flat spacetimes different techniques like
dimensional regularization and Green's function yield the same
results \cite{Milton:2001yy} this is not obvious for curved brane
world models. For instance the latter was adopted in
\cite{Linares:2007yz} for RSII-1 ($p$=1) in contrast with
\cite{Frank:2007jb} for RSII ($p$=0) based on dimensional
regularization and it would be desirable to be able to compare them.
Our analysis for $p$ arbitrary will allow to do so.

The interest in RSII-$p$ comes from its property of localizing not
only scalar and gravity but also gauge fields whenever $p>0$
\cite{Dubovsky:2000am,Dubovsky:2000av,Oda:2000zc}. It corresponds to
a $(3+p)$-brane with $p$ compact dimensions and positive tension
$\kappa$, embedded in a $(5+p)$-spacetime whose metric are two
patches of $AdS_{5+p}$ of curvature radius $\kappa^{-1}$
\begin{equation}\label{mebra}
ds_{5+p}^{\, 2}=e^{-2\kappa
|y|}\left[\eta_{\mu\nu}dx^{\mu}dx^{\nu}-\sum_{j\,=1}^pR^{\,
2}_jd\theta_j\right]-dy^2\,.
\end{equation}
Here $\eta_{\mu\nu}=$ diag$(+,-,-,-)$ is the 4D Minkowski metric,
$y$ is a coordinate for the non-compact dimension, $\theta_j$'s
$\in[0, 2\pi)$ with $j=0, \dots, p$, denote the coordinates for the
compact dimensions whereas $R_j$'s denote their size. Notice that
the $p$ compact dimensions are warped. This metric can be obtained
as an asymptotic solution to the $(5+p)$ Einstein equations with negative bulk
cosmological constant and a $(3+p)$-brane with an appropriately
tuned energy-momentum tensor
\cite{Gherghetta:2000qi,Gherghetta:2000jf}. It turns out that the
brane tension $\kappa$ is determined by the $(5+p)$-dimensional
Planck mass and cosmological constant.

We consider the standard setting of two parallel plates in 3-space
implemented by setting to zero  our scalar field on two planes in
ordinary 3-space (Dirichlet boundary conditions). As for the extra
dimensions we adopt conditions for the scalar field consistent with
RSII-$p$, namely matching across the brane for the non compact one
and periodic for the compact ones. Effectively then our ``{\em
plates}" are two parallel planes in ordinary three space stretching
along extra dimensions. The corresponding Casimir force hence
provides a generalization to brane worlds of that obtained by
Ambjorn and Wolfram \cite{Ambjorn:1981xw} for the case of arbitrary
number of Euclidean dimensions. Technically our plates are 2+1+$p$
dimensional embedded in an ambient space of dimension 3+1+$p$ so the
difference in dimension, or codimension, is 1. Similarly
\cite{Ambjorn:1981xw} considered $d$-1 dimensional plates embedded
in $d$ space.

Now as it was noticed in
\cite{Bajc:1999mh,Giddings:2000mu,Dubovsky:2000am,Linares:2007yz}
the scalar modes corresponding to the non compact dimension in
RSII-$p$ present different behavior depending on whether the scalar
field is massive or not. This justifies our splitting in two
separate sections for such two  cases.

This paper is organized as follows. Section II summarizes the
calculation of the vacuum force in RSII-$p$ based on Green's
function. Then in Sec. III the force is given  for the massive
scalar field whereas Sec. IV is devoted to the massless case.
Finally Sec. V contains the discussion of our results and further
developments. We use units in which $\hbar=1,c=1$.

\section{Casimir force and Green's function}

In this section we describe the ingredients and the strategy to get
the Casimir force for the scalar field within RSII-$p$ using the
Green's function. First we describe the origin of the scalar modes
associated to every dimension, ordinary or extra, by describing the
equations and boundary conditions they fulfill. They are then
incorporated to form the Green's function which is in turn related
to the vacuum energy momentum tensor of the scalar field. The
discontinuity of the latter for either of the plates provides then
the required vacuum force.

{\em Mode structure}. The starting point of our analysis is the
action in  $(5+p)$ dimensions for a scalar field $\Phi(X)$ in the
RSII-$p$ metric (\ref{mebra}), here denoted $g_{MN}$,
\begin{equation}\label{action}
S=\int \, d{}\,^4x  \,  \prod_{j=1}^p \, R_j d \theta_j \, dy\,
\sqrt{-g} \left( \frac{1}{2}\, g^{MN}\partial_M \Phi \,
\partial_N \Phi - \frac{1}{2} \, m_{5+p}^2 \, \Phi^2 \right)\, .
\end{equation}
Here $X^M\equiv(x^\mu,\theta_i, y)$ denote the associated
coordinates and $m_{(5+p)}$ is the mass of the field.  The
corresponding equation of motion for the scalar field is given
by
\begin{equation}\label{EqOfMotion}
e^{2\kappa|y|}\Box_4\Phi-e^{2\kappa|y|}\sum_{j=1}^p\frac{1}{R_j^{\,
2}}\,
\partial_{\theta_j}^{\, 2}\Phi -\frac{1}{\sqrt{-g}}\,
\partial_{y}\left[\sqrt{-g}
\partial_{y}\Phi\right]+m_{5+p}^2\Phi=0,
\end{equation}
which separates through $\Phi(X)=\varphi(x) \prod_{j=1}^p\, \Theta_j(\theta_j)\psi(y)$ into
\begin{eqnarray}
\left(\partial_{\theta_j}^{\, 2}+ m_{\theta_j}^{\,2}
R_j^{\,2}\right) \Theta_j (\theta_j)& = & 0,\qquad
j=1,\dots,p\label{Thetaeqn} \\
\left(\partial_{y}^{2}-(4+p)\kappa\,sgn(y)\partial_{y}-m_{5+p}^2+m^2\,e^{2\kappa|y|}\right)\psi(y)
& = & 0, \label{Yeqn}\\
\left(\Box_4 + \sum_{j=1}^p \, m_{\theta_j}^{\,2}+m^2 \right)\varphi
(x)& = & 0.\label{varphieqn}
\end{eqnarray}
The  $(p+1)$ separation constants with units of mass, $m_{\theta_j}$
and  $m$, correspond to the spectra of the modes for the compact and
non compact dimensions, respectively. They give rise in turn to the
effective mass, $m_4$,  of the 4D modes in (\ref{varphieqn}) through
\begin{equation}\label{m4} m_4^2:=\sum_{j=1}^p \,
m_{\theta_j}^{\,2}+m^2.
\end{equation}
To find mode solutions to the above Eqs. we require boundary
conditions that we next spell out.

{\em Boundary conditions}. We shall incorporate three types of
boundary conditions: (a) To implement the presence of the plates in
3-space we simply set $\varphi(z=0,l)=0$. (b) To match the modes
across the brane along the non compact dimension we impose
$\psi(y=0^{+})=\psi(y=0^{-})$ and
$\partial_y\psi(y=0^{+})=\partial_y\psi(y=0^{-})$. (c) To account
for the compactness of the $p$ dimensions we set
$\Theta_{n_j}(\theta_j)=\Theta_{n_j}(\theta_j+2\pi)$. Hereby we
obtain explicitly the plates represented by two parallel planes in
3-space and stretching along the extra dimensions. Conditions (a)
trough (c) are used for both cases $m_{5+p}\neq 0$ and $m_{5+p}=0$
which yield different spectra for the modes depending on the non
compact dimensions; they are denoted  $\psi_m(y)$, as described in
sect. III and IV. As for ordinary 3-space and $p$ compact dimensions
the result is identical for the two cases. The modes in $\theta_j$
are:
\begin{equation}\label{Theta}
\Theta_{n_j}(\theta_j)=\frac{1}{\sqrt{2\pi R_j}}e^{i\, n_j\,
\theta_j}\hspace{1cm} \mbox{where} \hspace{1cm} n_j=m_{\theta_j} R_j
\in {\mathbb{Z}}.
\end{equation}
Therefore the contributions of the extra compact dimensions to
$m_4$, are given in terms of $m_{\theta_j}^{\,2}=n_j^2/R_j^2$.
Rather than providing detailed modes for ordinary space its Green's
function fulfilling the appropriate boundary conditions is next
presented.

{\em Green's function}. We intend to determine the Green's function
for Eq. (\ref{EqOfMotion}) in terms of the modes (\ref{Thetaeqn})
and (\ref{Yeqn}) and the 4D Green's function corresponding to Eq.
(\ref{varphieqn}). As for the latter one considers separately the
two regions: between plates $(0<z<l)$ and, say, to the right of the
plate located at $z=l$. To do so let us split up the 3D position
vector in the way $\vec x =(\vec x_\perp,z)$, where $z$ denotes the
coordinate orthogonal to the plates and $\vec x_\perp$ denotes a 2D
vector orthogonal to the $z$ direction. Using this parametrization
the 4D Green's function is given by (see for instance
\cite{Milton:2001yy})
\begin{equation}\label{4DGreen}
G_{4D}\left(x,x';m_4^2\right)=\int
\frac{d\omega}{2\pi}\frac{d^2k_{\perp}}{(2\pi)^2} e^{-i\omega
(t-t')+i\vec{k}_{\perp}\cdot(\vec{x}_{\perp}-\vec{x}_{\perp}')}G(z,z')\,,
\end{equation}
where $m_4$ fulfills (\ref{m4}) and
\begin{equation}
G(z,z')=\begin{cases} G_{\mathrm{in}}(z,z')=-\frac{1}{\lambda \sin
\lambda \, l}\sin
\lambda \, z_< \sin \lambda(z_>-l),\quad 0\leq z,z' \leq l,\\
G_{\mathrm{out}}(z,z')=\frac{1}{\lambda}\left( \sin
{\lambda}(z_<-l)e^{i {\lambda}(z_>-l)} \right),\quad l\leq z,z' .
\end{cases}
\end{equation}
Here $z_>(z_<)$ represents the greater (lesser) of $z$ and $z'$ while the boundary conditions used between plates became
$ G(0,z')=G(l,z')=0$ and $G(z,z')\sim e^{ik z}$, as $z\to\infty$, to the right of the plate $z=l$.

The Green's function  for (\ref{EqOfMotion}) becomes \cite{Linares:2007yz}
\begin{equation}\label{6DComGreen}
G_{(5+p)D}=\sum_{\{n\}} \int^{\prime} dm\, \prod_{j=1}^p
\Theta_{n_j}^*(\theta_j) \Theta_{n_j} (\theta'_j)\psi_{
m}(y)\psi_{m}(y') G_{4D}\left(x,x';m_4^2\right),
\end{equation}
where $\{n\}$ denotes the set $\{n_1, n_2,\dots, n_p\, |\, n_1 \in
\mathbb{Z},\dots , n_p \in \mathbb{Z}\}$, the $\Theta_{n_j}$'s are
given by (\ref{Theta}),  the 4D Green's function by (\ref{4DGreen}),
and $dm$ is the measure for the continuous eigenvalues $m$ which to
include the zero value requires an extra term as it is seen below,
and hence the prime \cite{Randall:1999vf}. The $\psi_{m}$'s are
given in next two sections.

{\em Casimir force}. The knowledge of the Green's function allows to
obtain the force per unit area between the plates lying on ordinary
space as follows. Notice that the force between the plates is
obtained by integrating over coordinates ``lateral" to the plates.
In this case: $\vec{x}_{\perp},y,\theta_j$ due to the fact that the
normal-normal component of vacuum energy momentum tensor in 3+1+$p$
spatial dimensions has physical units of force per unit of ``volume"
of 2+1+$p$ space:
\begin{equation}\label{ForceT}
F = \int_0^A d\vec{x}_{\perp} \int_{-\infty}^{\infty} dy
\mathrm{e}^{-\kappa |y|(p+3)}\left[\prod_{j=1}^{p}\int_0^{2\pi}
Rd\theta_j\right] \left[\langle
T_{zz}^{\mathrm{in}}\rangle\bigg|_{z=l,y=0} - \langle
T_{zz}^{\mathrm{out}}\rangle\bigg|_{z=l,y=0}\right],
\end{equation}
where $A$ is the area of the planes forming the plates in 3-space so
we are taking a chunk of plate of volume
$A\left(\frac{2}{\kappa(3+p)}\right)(2\pi R)^p$, and we assume
$R_j=R$, $j=1,\dots,p$. As usual, $\langle\dots\rangle$ denotes
vacuum expectation values and the Green's function are related to
the normal-normal components of the vacuum energy momentum tensor
through
\begin{eqnarray}\label{t2}
\langle T_{zz}^{\mathrm{in/out}}\rangle\bigg|_{z=l,y=0} &=&
\frac{1}{2i}
\partial_z\partial_{z'}G_{(5+p)}^{\mathrm{in/out}}(x,y,\theta;x',y',\theta
')\bigg|_{x_\perp \rightarrow x_\perp', \, z\rightarrow z'=l,\,
y\rightarrow y'=0, \, \theta_j\rightarrow\theta_j '} \,.
\end{eqnarray}
The labels in/out in the r.h.s. of (\ref{t2}) imply the use of
(\ref{4DGreen}) in (\ref{6DComGreen}). To proceed further on the
details of the Casimir force we need to know the details of the
spectrum of the modes for the non compact dimension and hence we
separate the analysis depending on whether $m_{(5+p)}$ is zero or
not.

\section{Massive scalar field}
In this section we consider  $m_{5+p}\neq 0$. Subject to the above boundary conditions all modes are massive
\begin{equation}\label{psi}
\psi_m(y)=e^{\frac{4+p}{2}\kappa y}\,\sqrt{\frac{m}{2\kappa}}\left[
a_m J_{\gamma}\left(\frac{me^{\kappa y}}{\kappa}\right)+ b_m
N_{\gamma}\left(\frac{me^{\kappa y}}{\kappa}\right)  \right], \qquad m>0.
\end{equation}
$J_\gamma$ and $N_\gamma$ are the Bessel and Neumann
functions respectively. $\gamma=\sqrt{\left(\frac{4+p}{2}\right)^2+\left(\frac{m_{5+p}}{\kappa}\right)^2}$
and the coefficients $a_m$ and $b_m$ are given by
\begin{equation}\label{coeficients}
a_m=-\frac{A_m}{\sqrt{1+A_m^2}},\quad b_m=\frac{1}{\sqrt{1+A_m^2}},
\end{equation}
where
\begin{equation}\label{Aexpression}
A_m=\frac{N_{\gamma-1}\left(\frac{m}{\kappa}\right)-
\left(\gamma-\frac{4+p}{2}\right)\frac{\kappa}{m}N_{\gamma}\left(\frac{m}{\kappa}\right)}
{J_{\gamma-1}\left(\frac{m}{\kappa}\right)-
\left(\gamma-\frac{4+p}{2}\right)\frac{\kappa}{m}J_{\gamma}\left(\frac{m}{\kappa}\right)
}.
\end{equation}
At this point we can use the modes (\ref{psi}) into eqs.
(\ref{6DComGreen}), (\ref{ForceT}) and (\ref{t2}) to find the
Casimir force in the present case
\begin{eqnarray}
f_T:=\frac{F}{A}&=& \frac{2}{3+p}\sum_{\{n\}} \int
\frac{dm}{\kappa}\psi^2_{m}(0)\left(\frac{1}{2i}
\partial_z\partial_{z'}
G_{4D}^{\mathrm{in/out}}\left(x,x';m_4^2\right)\right)_{z\rightarrow
z'=l} \\
&=& \frac{2}{3+p} \sum_{\{n\}} \int \frac{dm}{\kappa}\psi^2_{m}(0)
f_{4D}(m_4)
\end{eqnarray}
since $f_{4D}(m_4)$ is the standard Casimir force for a scalar field of mass $m_4$
\begin{equation}\label{fCasimirStan}
f_{4D}(m_4)=-\frac{\hbar c}{32 \pi^2\, l^4} \, \int_{2lm_4}^{\infty}
dx \, \frac{x^2\sqrt{x^2-4\, l^{\, 2}\, m^2_4}}{e^x-1}\, ,
\end{equation}
with value given by (\ref{m4}), this result says extra dimensional
modes contribute individually as an admixture of massive modes
corresponding to compact and non compact dimensions but weighted by
the square of the wave function of the non compact modes at the
brane.

To extract more physical information some approximations are in
order. Although none of the modes (\ref{psi}) is localized on the
brane in the regime $m,m_{5+p}\ll \kappa$ there exists a
quasilocalized state corresponding to the complex eigenvalues $m=m_q
-i \Delta$ with
\begin{equation}\label{quasimass}
m_q^2=\frac{2+p}{4+p}\, m_{5+p}^2, \hspace{1cm} \mbox{and}
\hspace{1cm}\frac{\Delta}{m_q}=\frac{(2+p)\, \pi}{2^{\,
4+p}\left[\Gamma\left(\frac{4+p}{2}\right) \right]^2}\left(
\frac{m_q}{\kappa}\right)^{2+p}.
\end{equation}
A quasilocalized state can decay into the continuum modes.

At low energies, on the brane, there are two regions of the spectrum
which become relevant for which the small argument approximations
for Bessel and Neumann functions can be used
yielding
\begin{equation} \psi_m^2(0)\approx \left\{
\begin{array}{cc}\label{phiappoxn}
\frac{2+p}{2} \, \kappa   \, \delta(m_q-m) & \mbox{for}\, \, \, \, m \sim m_q ,  \\
\frac{2(2+p)}{\pi} \frac{\kappa \Delta}{m_q^{7+p}}\, m^{\, 5+p} &
\mbox{for}\, \, \, \, m\ll m_q .
\end{array} \right.
\end{equation}

In accordance with the low energy approximation we are using we can
assume $m\ll R^{-1}$ and therefore only the zero modes of the
compact coordinates are relevant ($n_1=\cdots=n_p=0$). The effective
4D Casimir force in this approximation is
\begin{eqnarray}\label{effectiveforce}
f_{\mathrm{eff}}&\approx& f_{\mathrm{quasi}}+f_{\mathrm{light}}=\frac{2+p}{3+p}f_{4D}(m_q)+f_{\mathrm{light}},\\
f_{\mathrm{light}}&:=&\frac{4(2+p)}{\pi (3+p)}
\frac{\Delta}{m_q^{7+p}} \int_{m\ll m_q} m^{\, 5+p} \, f_{4D}(m) \,
dm. \nonumber
\end{eqnarray}
The first term is proportional to the standard 4D Casimir force of a
massive scalar field whose mass is $m_q$, Eq. (\ref{quasimass}),
whereas the second term corresponds to the contribution of the light
modes. Upon the change of variables $m=\alpha/l$ we obtain
\begin{eqnarray}\label{LightContriMassive}
f_{\mathrm{light}} &\approx&
\frac{1920(2+p)}{\pi^3(3+p)}\frac{\Delta}{m_q}\frac{1}{(m_ql)^{6+p}}\,
f_{4D}(0)\, I(p),\\
I(p)&\equiv& \frac{1}{32 \pi^2} \, \int_0^{\infty} \alpha^{5+p}
\int_{2\alpha}^{\infty} dx \, \frac{x^2\sqrt{x^2-4\, \alpha^{\,
2}}}{e^x-1}\, ,\label{numericalintmassive}
\end{eqnarray}
where $f_{4D}(0)=-\frac{\pi^2}{480}\frac {\hbar c}{l^4}$ represents
the standard Casimir force of a massless scalar field. The numerical values for $I(p)$
are tabulated in table \ref{massive}, for $p$ in the range $0,
\dots, 4$.
\begin{table}
\begin{tabular}{||c||c|c|c|c|c||}\hline
$p$ & 0 & 1 & 2 & 3 & 4 \\ \hline $I(p)$ & 1.37 & 5.51 & 25.08 & 127.25 & 711.32\\
\hline
\end{tabular}
\caption{Numerical value of the integral (\ref{numericalintmassive})
for different number of extra compact dimensions $p$.}
\label{massive}
\end{table}

Recalling the light scalar field approximation $m_ql\ll 1
\Rightarrow f_{4D}(m_q)\approx f_{4D}(0)$ and considering
$f_{\mathrm{light}}\ll \frac{2+p}{3+p} f_{4D}(0)$ in Eq.
(\ref{effectiveforce}) one gets the lower bound $m_q \, (m_q
l)^{6+p}/\Delta \gg (1920)I(p)/\pi^3$, or equivalently, $(\kappa
l)^{2+p}(m_q l)^4 \gg 120 (2+p)I(p)/\pi^2 2^p
\,[\Gamma((4+p)/2)]^2$.

\section{Massless scalar field}
Now we use $m_{5+p}=0$. The allowed modes for the non compact
dimension are now a massless zero mode localized on the
brane
\begin{equation}\label{zeromode}
\psi_0=\sqrt{\frac{(2+p)\kappa}{2}},
\end{equation}
which satisfies the normalization condition,
$\int_{-\infty}^{\infty} \, dy \,  e^{-2\kappa |y|}\, \psi_0^2 $ and
the massive modes have the form (\ref{psi}) by taking $m_{5+p}=0$
thus getting $\gamma\rightarrow \gamma_0=\frac{4+p}{2}$ and
$A_m\rightarrow
A^0_m=\frac{N_{\gamma_0-1}\left(\frac{m}{\kappa}\right)}{
J_{\gamma_0-1}\left(\frac{m}{\kappa}\right)}$.

Here the Casimir force, upon use of these modes, becomes
\begin{equation}
f_T = \frac{2}{3+p}\sum_{\{n\}} \left[ \frac{\psi_0^2}{\kappa} \,
f_{4D}\left( m_4\bigg|_{m=0}\right)+\int_0^{\infty}\frac{dm}{\kappa}
\psi_m^2(0) f_{4D}\left(m_4\right) \right],
\end{equation}
which can be read as saying that extra dimensional modes contribute
to the Casimir force as individual discrete modes or as an admixture
of discrete and continuous modes weighted by squared wave functions
of the modes in the non compact direction on the brane.

In the low energy regime $m\ll \kappa$ and only the zero modes
associated to the compact extra dimensions are relevant
($n_1=\dots=n_p=0$) so
\begin{equation}
A^0_m \approx -\frac{\Gamma(\gamma_0 )\Gamma(\gamma_0
-1)}{\pi}\left(\frac{m}{2\kappa}\right)^{2-2\gamma},
\end{equation}
and the wave function of the continuous modes at the brane becomes
\begin{equation}
\psi_m^2(y\rightarrow 0) \approx \frac{1}{2^{p+1}\left[ \Gamma\left(
\frac{2+p}{2}\right)\right]^2} \frac{m^{p+1}}{\kappa^{p+1}}.
\end{equation}
The Casimir force takes the form
\begin{eqnarray}\label{LightMassless}
f^0_{\mathrm{eff}}&\approx&  \frac{2+p}{3+p} f_{4D}(m_4=0)+f^0_{\mathrm{light}}, \\
f^0_{\mathrm{light}} &:=& \frac{1}{2^{\, p}\, (3+p)
\left[\Gamma\left(\frac{2+p}{2}\right) \right]^2}
\frac{1}{\kappa^{\, 2+p}} \int_{m\ll \kappa} m^{\, 1+p} \, f_{4D}(m)
\, dm, \nonumber
\end{eqnarray}
with $f_{4D}(m)$ given by (\ref{fCasimirStan}). The light modes
contribution can be evaluated further as
\begin{equation}\label{LightContriMassless}
f^0_{\mathrm{light}} \approx \frac{480}{\pi^2\, 2^{\, p}\, (3+p)
\left[\Gamma\left(\frac{2+p}{2}\right) \right]^2} \frac{1}{(\kappa
l)^{\, 2+p}} \, f_{4D}(0)\, J(p),
\end{equation}
where $J(p)$ is defined by
\begin{equation}\label{numericalintmassless}
J(p)\equiv \frac{1}{32 \pi^2} \, \int_0^{\infty} \alpha^{1+p}
\int_{2\alpha}^{\infty} dx \, \frac{x^2\sqrt{x^2-4\, \alpha^{\,
2}}}{e^x-1}\, .
\end{equation}
In table \ref{massless} we give
the value of this integral for $p$ in the range $0, \dots, 4$.
\begin{table}
\begin{tabular}{||c||c|c|c|c|c||}\hline
$p$ & 0 & 1 & 2 & 3 & 4 \\ \hline
$J(p)$ & 0.032 & 0.056 & 0.134 & 0.392 & 1.369 \\
\hline
\end{tabular}
\caption{Numerical value of the integral
(\ref{numericalintmassless}) for different number of extra compact
dimensions $p$.} \label{massless}
\end{table}

Considering again $f_{\mathrm{light}}\ll \frac{2+p}{3+p} f_{4D}(0)$
in (\ref{LightMassless}) one gets the lower bound $(\kappa l)^{2+p}
\gg 480 \, J(p) /3 \, \pi^2\, 2^{\, p}\,
\left[\Gamma\left(\frac{2+p}{2}\right)\right]^2$. By taking $l \sim
10^{-6}$m of typical Casimir experiments one gets an upper bound for
the anti de Sitter radius of $\kappa^{-1}\ll (3 \, \pi^2\, 2^{\,
p}\, \left[\Gamma\left(\frac{2+p}{2}\right)\right]^2/480 \,
J(p))^{1/(2+p)} \times 10^{-6}$m.

\section{Discussion}

Casimir or vacuum force
\cite{Casimir:1948dh,Sparnaay:1958wg,Mohideen:1998iz,Bressi:2002fr,Decca:2005yk,Klimchitskaya:2005df,Lamoreaux:2005gf,Milton:2001yy}
appears amongst a set of little explored low energy tests, possibly
including high precision atomic experiments \cite{Bluhm:2000tv},
amenable to probe brane worlds. Two of the reasons to consider it
are the fact it depends on the mode structure of matter fields and
also that experimentally it lies in the range of recent and future
sub-millimeter gravity experiments
\cite{Hoyle:2000cv,Adelberger:2006dh,Kapner:2006si}, both of which
are related to brane worlds models.  With this motivation in this
letter we have determined brane world corrections to standard 4D
Casimir force for a scalar field subject to Dirichlet boundary
conditions on two parallel plates lying within the single
(3+$p$)-brane of Randall-Sundrum type of scenario or RSII-$p$. We
adopt the Green's function approach to do so and hence as a first
step a study of the modes for the scalar field was developed in the
background brane world.

The resulting scalar modes group themselves as a continuous tower of
massive modes presenting a single quasilocalized massive mode or as
a massless mode incorporated with a continuous tower of massive
modes depending on whether the scalar field is massive or not. In
the low energy approximation for either case only the zero modes
corresponding to the compact dimensions and light modes for the non
compact dimension are considered. They yield a Casimir force
naturally splitting into a leading order term given by the
quasilocalized or the massless mode, for the massive and massless
scalar field, respectively, plus a correction term coming from the
light continuous modes, Eqs. (\ref{effectiveforce}) and
(\ref{LightMassless}). Such leading order terms coincide with the
standard 4D Casimir force for massive or massless scalar field,
respectively, up to numerical factors depending on the number of
compact dimensions $p$. Moreover corrections turn out to be
attractive and depend on the separation between plates as
$l^{-(10+p)}$ for the massive scalar field and as $l^{-(6+p)}$ for
the massless one, Eqs. (\ref{LightContriMassive}) and
(\ref{LightContriMassless}), respectively. To keep corrections
negligible compared to the 4D force upper bound for $AdS_{(5+p)}$
radius $\kappa^{-1}$ is obtained of $\sim 10^{-6}$m.

It is illustrative to compare the obtained Casimir force with others
in recent literature. For $p=1$ the results of \cite{Linares:2007yz}
are recovered except for a numerical factor: $\frac{3}{4}$ instead
of $\frac{3}{2}$ in Eqs. (\ref{effectiveforce}) and
(\ref{LightMassless}) due to the volume factor (\ref{ForceT}).
Furthermore, the case $p=0$ is of particular interest since it was
studied in \cite{Frank:2007jb} by the dimensional regularization
technique. It is worth stressing that this case fails to be an adequate model
for the realistic electromagnetic case since for the latter localization does not work as opposed to the scalar field.
Still, theoretically, its study sheds light on the comparison between different approaches. In \cite{Frank:2007jb} they get
\begin{equation}\label{discordia1}
F_{Total}=f_{4D}(0)\left( 1 + \frac{45}{2\pi^3}\zeta(5)\frac{1}{l
\kappa}\right),
\end{equation}
which clearly differs from ours, Eqs.
(\ref{LightMassless})-(\ref{LightContriMassless}) for $p=0$. We now
comment on this difference. It was shown in \cite{Ambjorn:1981xw}
that the Casimir force per unit area on the boundary plates for a
scalar field in a spacetime with $d$ Euclidean spatial dimensions is
always attractive and is given by
\begin{equation}\label{forceAmbjorn}
f_d=
-\frac{d}{l^{d+1}}\Gamma\left(\frac{d+1}{2}\right)(4\pi)^{-(d+1)/2}
\zeta(d+1).
\end{equation}
If we sum the Casimir force for a massless scalar field in 4D
($d$=3) with the Casimir force for a massless scalar field in  5D
($d$=4) we obtain
\begin{equation}\label{discordia2}
F=f_{4D}\left( 1 + \frac{45}{4\pi^4}\zeta(5)\frac{1}{l}\right).
\end{equation}
Comparing (\ref{discordia1}) with (\ref{discordia2}) we see that
these equations differ only by a factor $2\pi /\kappa$ in the second
term which can be understood as follows. The modified dispersion
relations used in obtaining (\ref{discordia1}) have an extra
continuous mass term  $m^2$ with $m \in [0,\infty)$, therefore when
computing the force there is a term with an extra integral $\int
dm/\kappa= (2\pi /\kappa)\int dm /2\pi$, which amounts to have
$(2\pi /\kappa)f_{d=4}$ that corresponds to $2\pi /\kappa$ times the
force for a 5D scalar field in flat space. Interpreting
(\ref{discordia1}) now is clear: it contains two terms, the standard
4D Casimir force for a massless scalar field in Minkowski spacetime
plus $2\pi /\kappa$ times the Casimir force for a 5D massless scalar
field in 5D Minkowski spacetime. Indeed dimensional regularization as
considered above is blind to physical information like the curvature
of the background. Such state of affairs could be remedied by considering
dimensional regularization on curved backgrounds \cite{Elizalde:2006iu}.
Then both approaches could be fairly compared.

Future problems suggest themselves based on the previous analysis.
For instance, among several other models, it would be interesting to
obtain the effective Casimir force for a scalar field in the two
branes RSI background metric, possible extended by compact
dimensions, using Green's function method to contrast it with that
obtained by dimensional regularization in \cite{Frank:2007jb}.

Finally, a comment regarding boundary conditions is in order. The
scalar modes have been built to fulfil Dirichlet boundary conditions
in ordinary 3-space, to implement the presence of the plates here
represented by two planes. Along the extra dimensions the boundary
conditions used were those appropriate for the background brane
model: matching along the brane for the non compact dimension and
periodic for the compact ones. Hence, as stressed in this work, we
have plates extended along extra dimensions which, however -and this
is crucial physically- appear as two parallel planes within ordinary
3-space. Indeed our Casimir setting works as shown by the leading
order terms corresponding to those in 4D, Eqs.
(\ref{effectiveforce}) and (\ref{LightMassless}). Moreover it
generalizes to brane worlds of type RSII-$p$ previous results for
arbitrary number of flat dimensions with two codimension one plates
\cite{Ambjorn:1981xw}. It would be interesting to study the
situation in which one restrains the plates to 3-space to coincide
with two planes without stretching along extra dimensions. They
would have dimension 2 within 3+1+$p$ space, thus having codimension
2+$p$ (see for instance \cite{Scardicchio:2005hh}).

\begin{acknowledgments}
This work was partially supported by Mexico's National Council of
Science and Technology (CONACyT), under grants
CONACyT-SEP-2004-C01-47597, CONACyT-SEP-2005-C01-51132F and a
sabbatical grant to HAMT. The work of O.P. was supported by a
scholarship of the grant CONACyT-SEP-2004-C01-47597.
\end{acknowledgments}
%\newpage %Just because of unusual number of tables stacked at end
\bibliography{bibliography}% Produces the bibliography via BibTeX.

\end{document}